\documentclass{aastex}\usepackage{emulateapj5}
\usepackage{rotating,epsf}
\def\msun{\mbox{ M}_\odot}
\def\mag{\mbox{ mag}}

\lefthead{Metcalf \& Zhao}
\righthead{Flux ratios as a probe of dark substructures}

\begin{document}
%\special{!userdict begin /bop-hook{gsave 200 30 translate 65 rotate 
%/Times-Roman findfont 220 scalefont setfont 0 0 moveto 0.93 setgray 
%(DRAFT) show grestore}def end}

\title{Flux ratios as a probe of dark substructures
in quadruple-image gravitational lenses}
\author{R. Benton Metcalf\altaffilmark{1} \& HongSheng Zhao\altaffilmark{1}}
\affil{Institute of Astronomy, Cambridge, CB3 0HA, UK}
\altaffiltext{1}{Email: {\tt bmetcalf@ast.cam.ac.uk, zhao@ast.cam.ac.uk}}

%\maketitle

\begin{abstract}
We demonstrate that the flux ratios of 4-image lensed quasars  
provide a powerful means of probing the small scale structure of 
Dark Matter (DM) halos.  A family of smooth lens models can precisely
predict certain combinations of flux ratios using only the positions of
the images and lens as inputs.  Using 5 observed lens systems we show
that real galaxies {\it cannot} be described by smooth 
singular isothermal ellipsoids, nor by the more general
elliptical power-law potentials.
Large scale distortions from the elliptical models can not yet be ruled out.
Nevertheless we find, by comparing with simulations, that the data can
be accounted for by a significant ($\ge 5-10\%$) 
amount of dark substructures within 
a projected distance of several kpc from the center of lenses.
This interpretation favors the Cold Dark Matter (CDM) model other
the warm or self-interacting DM models.

\end{abstract}


\section{Introduction}
  \label{sec:intro}

The standard $\Lambda$CDM cosmological model has been very
successful in accounting for observations on scales larger than  
around a Mpc.  But mounting evidence points towards
difficulties for this theory on the scales of galaxies and dwarf
galaxies \cite{vdB:00,GZ01}.
Most interestingly for this paper, CDM
simulations of the local group of galaxies predict an order of magnitude
more dwarf galaxy halos with masses greater than $\sim 10^8\,\msun$ than
there are observed satellites of the Milky Way (MW) Galaxy and the M31
galaxy \cite{1999ApJ...524L..19M,1999ApJ...522...82K,Mateo}.  

This overprediction of dwarf halos could be a sign that there is
something fundamentally wrong with the CDM model
-- the variants include warm dark matter
(e.g., Bode, Ostriker \& Turok 2001), self-interacting dark matter
\cite{SS:00} and unorthodox inflation models \cite{2000PRL.Kamionkowski}.  Alternatively, the small DM clumps
could exist, but not contain stars, so as to escape detection as
observable dwarf galaxies.  This situation can easily, perhaps
inevitably, come about through the action of feedback processes in the
early universe (photoionization and supernova winds)
(e.g., Somerville 2001).
%For example, photoionization can prevent gas from
%cooling and thus inhibit star formation in small halos
Several authors, e.g. Metcalf (2001),
have argue that the overabundance of DM clumps is likely to extend down
to smaller masses and larger fractions of the halo mass than have thus
far been accessible to numerical simulations.
These nearly pure dark matter structures have largely been considered
undetectable.

Metcalf \& Madau (2001) showed that if the CDM model is correct and
these substructures exist within the strong gravitational lenses 
responsible for multiply imaged QSOs they will have a dramatic effect on
the image magnifications -- compound gravitational lensing.  It was
proposed that the image positions which are only weakly affected by
substructure could be used to constrain a smooth model for the lens.
The magnification ratios can then be compared with the predictions of this
model.  However, to detect this effect the model or family of models
must accurately predict the magnification ratios at around the $\sim
0.1\mag$ level.  It must be certain that there is no smooth lens model
which can reproduce both the observed ratios and image positions.  This is
the task that is taken up in this paper.

Mao \& Schnieder (1998) first proposed that the magnification
ratios of B1422+231 are better fit by a model with substructure in it
than with a smooth model.  They fixed the smooth model and added point
masses to represent globular clusters and plane wave perturbations.
They found that they could reproduce the ratios with small scale
fluctuations of relatively large amplitude.  This does not however
explain the discrepancies between radio and optical ratios in this
system.  Recently Chiba (2001) has done a similar analysis where
the smooth model is fixed and masses are added.  He concludes that
globular clusters and dwarf galaxies (as represented by point masses)
are not sufficient to reproduce the data.  Both these investigations
were restricted to the singular isothermal ellipsoidal models for 
the lens galaxy, an assumption which we do not make here.
During the refereeing process of this paper a few other preprints appeared
on very similar topics (Dalal \& Kochanek 2001, Keeton 2001 and 
Brada\v{c} {\it et al.} 2001).  

\section{Method}
  \label{sec:method}
\subsection{The Lens model}

The lensing equation, $\vec{z}=\vec{x}-\vec{\alpha}(\vec{x})$, relates a
point on the source plane $\vec{z}$ to a point on the image plain $\vec{x}$,
where $\vec{\alpha}(\vec{x})$ is the deflection angle.  These are
angular positions.  The deflection angle can be expressed as the
gradient of a lensing potential: $\vec{\alpha}(\vec{x})=\vec{\nabla}\psi(\vec{x})$.
This potential is related to the surface density of the lens,
$\Sigma(\vec{x})$, through the two dimensional Poisson equation
\begin{equation}
\nabla^2\psi(\vec{x})=2\kappa(\vec{x})~~,~~\kappa(\vec{x})=\Sigma(\vec{x})/\Sigma_c.
\end{equation}
where $\Sigma_c\equiv c^2 D_s(4\pi G D_lD_{\rm ls})^{-1}$ is the
critical surface density.  Here $D_l$, $D_s$, and $D_{\rm ls}$ are the
angular size distances to the lens, source, and from the lens to the source, respectively.

To model the lens galaxy we use an elliptical potential with a power-law
radial profile (EPL).  The influence of other galaxies that might
neighbor the primary lens is included through a background shear
(EPL+S).  The potential for these models is

\begin{eqnarray} \label{potential}
\psi(x,y)= b r_\epsilon^{2n} 
%+ \frac{1}{2}\gamma r^2\cos2(\theta-\theta_\gamma) 
+ \frac{1}{2}\gamma \left[ (\Delta x^2 -\Delta y^2)\cos
2\theta_\gamma \right.\\  \nonumber
\left.+ 2\Delta x\Delta y\sin 2\theta_\gamma \right]
\end{eqnarray}
\begin{eqnarray}
\begin{array}{c}
\Delta x \equiv (x-x_l)~~,~~\Delta y \equiv (y-y_l)~~~~~~~~ \\
r_\epsilon^2 \equiv ~~\Delta x^2(\cos^2\theta_\epsilon + \epsilon^2
\sin^2\theta_\epsilon) ~~~~~~~~~~\\+ \Delta y^2(\sin^2\theta_\epsilon + \epsilon^2
\cos^2\theta_\epsilon) \\+ \Delta x \Delta y(1-\epsilon^2)
\sin2\theta_\epsilon~~~~
\end{array}
\end{eqnarray}
where the center of the lens is located at $\vec{x}_l=(x_l,y_l)$.  The
second term in the potential (\ref{potential}) is the external shear.
With the addition of the source position the number of free parameters
comes to 10.  A singular isothermal sphere corresponds to $n=0.5$ and
$\epsilon=1$ and a point mass would be indicated if $n$ were close to
zero (in this case $\psi\propto \ln r$).

\subsection{Searching parameter space}

Typically the lens model and thus the likelihood are functions of many
parameters -- in our case 10.  As a result characterizing the
properties of a family of models in any more detail then simply finding
the best fitting set of parameters can be difficult.  Most, perhaps all,
authors have taken a Monte Carlo approach where the image positions are
chosen at random according to the observational uncertainties.  Then the
best--fit model is found and the parameters of that model are recorded.
This is {\it not} the correct procedure.  There is not generally a
one-to-one correspondence between image positions and model parameters;
not all sets of image positions correspond exactly to a set of model
parameters.  The region over which one can legally vary the positions is
highly restricted (in 10 dimensions).  In addition, if the model has a
high degree of freedom multiple sets of parameters may correspond to the
same image positions.

To avoid these problems we generate random models and then calculate the
image positions.  This is significantly more time consuming because one
does not know how well a parameter set fits the data until all the
calculations are finished.  To achieve any degree of efficiency and to
ferret out the corners of confidence regions some adaptive sampling of
the parameters must be used.  We do this by first finding the best--fit
model and set a maximum $\chi^2$ that we consider an acceptable fit.  
The models are chosen according to the probability $P({\bf x})=\prod_i
p(x_i,\sigma_i)$ where each $p(x_i,\sigma_i)$ is a normal
distribution.  The parameters $x_i$ are the distance from the best--fit
model along each of the principle directions of the 
distribution of models already found in the acceptable--fit region.
These directions are periodically updated.  The standard deviations
$\sigma_i$ are updated so that it is some fixed factor times the distance to
most extreme acceptable--fit model in that direction.  This factor is
increased until the final confidence region is no longer dependent on
it, typically $\sim\sqrt{3}$.  This algorithm works well except when the
acceptable--fit region is highly curved in parameter space or has long
thin regions projecting from a thicker region.
The most difficult task is to find the boundaries of the very
high confidence regions.  As a result the 95\% confidence intervals
reported here are more secure than the 99\% ones although we do believe
that the calculation has converged in all cases.

In this paper we are primarily interested in the magnification ratios.
For a 4-image system we can combine the fluxes into three
independent quantities of the form
\begin{equation}
q^j=\sum_{i=1}^4 a_i^j m^i ~;~ \sum_{i=1}^4 a_i^j = 0
\end{equation}
where $m^i$ is the magnitude of the $i$th image.  The constraint on
$a_i^j$ ensures that $q^j$ is independent of the source luminosity.
In addition we require that the transformation from the observed
magnitudes be orthogonal so that the $q^i$'s
are still measured in magnitudes.  We choose three orthogonal
generalized flux ratios such that one of them has the
smallest confidence interval and one of them has the largest.  This
displays the maximum precision to which the magnification ratios can be
predicted.

\section{Data}

We model five 4-image systems (MG0414+0534, B1422+231, PG1115+080,
Q2237+030 and H1413+117) using publicly available data.  For the
positions of the QSO images and the center of the lens galaxy we use HST
data available on the CASTLES Survey's web site
(http://cfa-www.harvard.edu/castles/).  We use the infrared and visible
extinction-corrected flux ratios from Falco {\it et al.} (1999).  The errors in the
IR/visible ratios are dominated by uncertainties in the extinction
correction rather than photometric errors.  The radio data
comes from Katz, Moore \& Hewitt (1997) for MG0414, from Patnaik {\it el
al.} (1992) for B1422, and from Falco {\it el al.} (1996) for Q2237.

\section{Results}
\label{sec:results}

\begin{figure}
\epsfxsize=10cm
\epsfbox{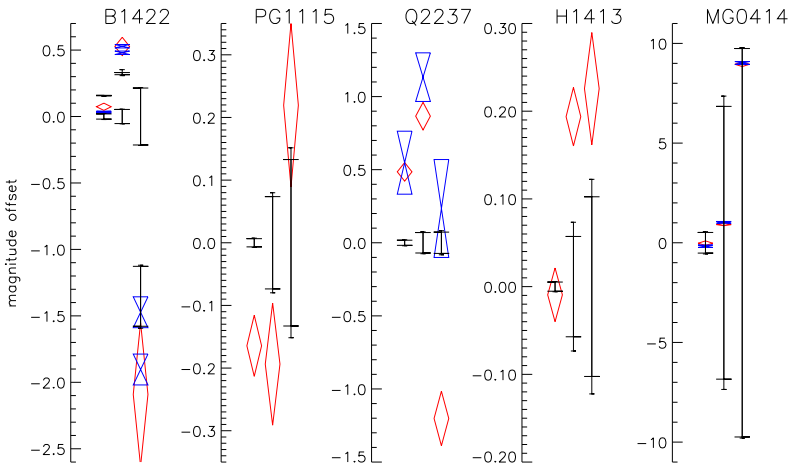}
\caption{\footnotesize The confidence ranges for the EPL+S in the optimal three
independent combinations of the magnification ratios for five lens
systems.  The 95\% and 99\% confidence regions are shown by the
two sets of horizontal bars on each error bar.  
The bow ties mark the radio ratios
at 5~GHz except for B1422 where both 5~GHz and 8~GHz ratios are shown.
The size of the bow ties are the reported errors.
The diamonds mark the extinction--corrected infrared and visible ratios
with the lengths giving the errors \cite{Falco99}.
}
\label{ranges}

\end{figure}

In Table~\ref{table} some of the best--fit lens parameters for the
five 4-image systems are given along with the coefficients for the
generalized magnification ratio.  It should be noted that in 
practice the likelihood is often rather flat in some directions near the
maximum as is most clearly the case in MG0414 (see figure~\ref{ranges}).
Our Monte Carlo technique tries to account for these difficulties. 
In addition there can also be local maxima that are well separated in parameter
space.  For B1422 we find two well fitting regions of parameter
space.  They do not appear to be connected at any reasonable confidence
level.  In every case the best--fit model fits the image and lens positions
very well which is expected since we have an equal number of constraints
and parameters.  The image positions are all consistent with the
hypothesis that the EPL+S models correctly describe the lens potentials.

%Occasionally the substructure should shift the image positions on the
%scale of the observational errors (Metcalf \& Madau 2001).

Figure~\ref{ranges} shows the magnification ratio confidence regions
along with the data.  The generalized magnification ratios have been
shifted for convenience so that the confidence region is centered on
zero magnitudes in all cases except B1422 where we have two acceptable--fit
regions.  Here the ratios are shifted so that the region with the
globally best fitting solution is centered on 0.   It is significant
that the width 99\% confidence interval for the best generalized ratio is
$<0.06$~mag in all cases except MG0414 where it is 1.1~mag.  The data is also
plotted in figure~\ref{ranges} with the size of the symbols representing
the errors transformed into the generalized ratios.  One can immediately see that
many of the measurements are inconsistent with the smooth EPL+S model.
Only MG0414 is fully consistent with the models.  Of the two acceptable
regions for B1422 it is the one that fits the image positions less well
that fits the flux ratios best.

%Most striking are the radio
%measurements of B1422, the visible/IR observations of H1413, and the
%observations of Q2237 in both bands. 

\section{Lensing Simulations}

For comparison we have done some simulations of what is expected
in the CDM model.  We have not attempted to account for the fact that the
image positions do not constrain the smooth model to a precise set of
smooth model parameters.  Because the effects of substructure are strongly
dependent on the smooth model used only a qualitative comparison with the
CDM model is now possible.  We will make a more systematic study of this
problem in future work.

The code used to do these simulations is described in
Metcalf \& Madau (2001).  The smooth models are fixed to the best fit
models given in Table~\ref{table}.  The substructure is modeled as
singular isothermal spheres cut off at the tidal radius.  The mass
function is $\propto m^{-2}$ between $10^4\msun$ and $10^6\msun$ and the
normalization is fixed so that 5\% and 10\% of the lens surface density
is contained in the substructure.  The physical radius of the sources are set
to 30~pc which is perhaps appropriate for the radio emission.  We
calculate 1500 random realization of the ratios for each lens.  We did
not simulate H1413 because the lens redshift is not known

\begin{figure}
\epsfxsize=10cm
\epsfbox{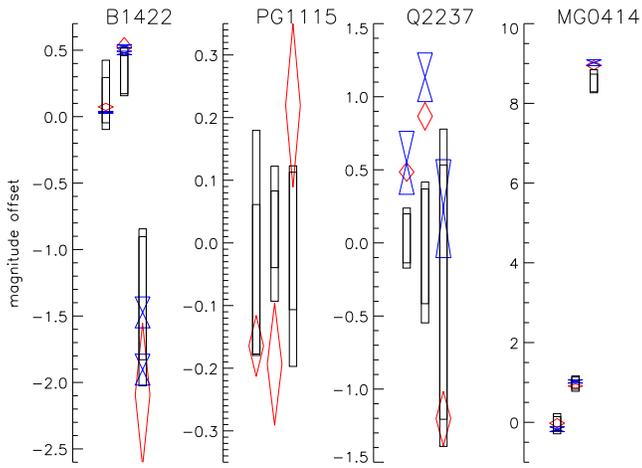}
\caption{\footnotesize Same as figure~\ref{ranges} except that the
smooth models are fixed and the range of magnification ratios expected
in the presence of substructure are shown as rectangles.  The rectangles
show the region in which 95\% of the simulated realizations fall.  There are
two rectangles for each ratio, one inside the other.  The smaller ones
are for the case of 5\% of the surface density in substructure and the
larger ones are for 10\%.  The scale has been changed from
figure~\ref{ranges} for MG0414 for clarity.
}
\label{compound_ranges}
\end{figure}

The results are shown in figure~\ref{compound_ranges}.  Except for the
second ratio of Q2237, all of the anomalous ratios fall
within 2 $\sigma$ of the regions containing 95\% of the simulated ratios
if 10\% of the mass is put in substructure.  If anything the data
appears to favor even more substructure or higher mass substructure
relative to the source size.

\section{Discussion}
  \label{sec:discussion}

In all cases except MG0414 the EPL+S models are ruled out at greater than
95\% confidence.  The radio fluxes are not affected by extinction or
microlensing and they alone rule out the models for B1422 and Q2237.  
The disagreement between visible and radio ratios in Q2237 is 
confirmation of the microlensing that was already known to exist 
through time variability studies \cite[and references there
in]{OGLE_Q2237}.  The strong disagreement between 
the radio and model predictions for Q2237 does indicate that there is some
other kind of substructure there as well.  Given the absence of
microlensing-related strong variability in PG1115 or H1413 the observed
visible/IR flux ratios appear inconsistent with EPL+S models.  
This is evidence that a significant amount of small scale structure must
exist either in the lenses or along the lines of sight.  

The known populations of small scale substructures in the Galaxy would be 
unlikely to cause the effects reported here.  The overdensities in spiral arms do
not appear large enough and, as pointed out by Mao \&
Schneider (1998), the fraction of the Galaxy halo's mass in globular
clusters is only about $\sim 10^{-4}$.  The mass in dwarf galaxy
satellites is $\sim 1\%$ of the halo, but $80- 90\%$ of this is in just two
objects.  One would expect a chance alignment of these types of
structures with the QSO images to be rare not ubiquitous.

Within the EPL+S models the magnification ratios are
quite strongly constrained in some dimensions.  In all but MG0414 we were
able to constrain one combination of magnification ratios to within
0.06~mag.  Within a large family of
galactic mass profiles the ratios can be an effective tool for probing
small scale structure.  In simulations with pure CDM, galaxies do have power-law
radial profiles within the small radial distances important for
quad-lenses.  And observed lens galaxies are typically relaxed giant
ellipticals.  However, there is no compelling reason to believe that 
CDM halos and their embedded galaxies should be precisely elliptical;
bulges and inclined disks could make the lens mass distribution more
boxy or more disky in projection in the inner few kpc.  Zhao \& Pronk
(2001) have found that lens models with different degrees of boxiness 
can fit the image positions equally well, but predict different flux ratios
and time delay ratios.  This leads to an ambiguity in our
substructure interpretation with large scale distortion.  
Ultimately this degeneracy can be lifted
by either restricting the range of realistic halo profiles using
simulations or by comparing observations in different wavelengths -- the
magnifications should be wavelength dependent (Metcalf \& Madau 2001).

The constraints on the predicted magnification ratios are good, but they are not
negligible.  In some cases the uncertainty in the ratios is much larger
than the expected variations caused by substructure.  The influence of
substructure on the ratios can be strongly dependent on the precise
values of smooth model parameters.  This should be taken into account in
future compound lensing studies.

This result is in rough agreement with what is expected in the
CDM model (Metcalf \& Madau 2001) as shown by our simulations.  In
variants of the CDM model such as warm DM, hot DM or interacting DM
small scale substructure is almost nonexistent.  These models are not
simultaneously consistent with the EPL+S lens models and the observed
flux ratios.  Possible degeneracies with the larger scale distortions of
the lens make us unable to conclusively discriminate between DM models
at present.

\bigskip\noindent 
{\bf Acknowledgements} \\ 
RBM would like to thank Piero Madau for very useful conversations.

\onecolumn

% \begin{table}
% \begin{tabular}{lccccccccc}
% & \multicolumn{5}{c}{Best--Fit Lens Models} &~&
% \multicolumn{3}{c}{Generalized Ratios} \\
% \multicolumn{1}{c}{Lens} & $P(\chi^2)$ & $b$ &
% $\gamma$ & $\epsilon^2$ & $n$ &~& $a_1$ & $a_2$ & $a_3$ \\
% B1422+231   & 0.00 & 0.91 & 0.25 & 1.58 & 0.32 &~& -0.62 & 0.78 & -0.14 \\
%             & 0.00 & 2.00 & 0.36 & 2.38 & 0.11 &~& -0.73 & 0.69 & -0.03 \\
% PG1115+080  & 0.00 & 14.7 & 0.21 & 1.13 & 0.04 &~& -0.76 & 0.65 &  0.08 \\
% Q2237+030   & 0.00 & 1.44 & 0.05 & 1.37 & 0.28 &~&  0.77 & 0.01 & -0.63 \\
% H1413+117   & 0.21 & 0.45 & 0.10 & 1.21 & 0.97 &~&  0.32 & 0.44 & -0.84 \\
% MG0414+0534 & 0.00 & 0.10 & 0.85 & 0.52 & 0.65 &~& -0.004&-0.01 &  -0.999
% \end{tabular}
% \caption{\footnotesize The $a_i$ coefficients are for the most well constrained
% generalized magnification ratio ($a_4=-\sum_{i=1}^3 a_i$).  The
% numbering of images is in the same order as those in the observational
% liturature.  $P(\chi^2)$ is the expected probability (rounded off) of
% $\chi^2$ being smaller than the one measured with 10 degrees of
% freedom.  Of the two models for B1422 the second one is a slightly
% better fit to the image positions.
% \label{table}}
% \end{table}

\begin{table}
\begin{tabular}{lccccccccc}
& \multicolumn{5}{c}{Best--Fit Lens Models} &~&
\multicolumn{3}{c}{Generalized Ratios} \\
\multicolumn{1}{c}{Lens} & $P(\chi^2)$ & $b$ &
$\gamma$ & $\epsilon^2$ & $n$ &~& $a_1$ & $a_2$ & $a_3$ \\
B1422+231   & 0.00 & 0.91 & 0.25 & 1.58 & 0.32 &~& -0.61 & 0.76 & -0.20 \\
            &      &      &      &      &      &~&  0.61 & 0.62 &  0.49 \\
            & 0.00 & 2.00 & 0.36 & 2.38 & 0.11 &~& -0.73 & 0.69 & -0.03 \\
            &      &      &      &      &      &~&  0.61 & 0.62 &  0.49 \\
PG1115+080  & 0.00 & 14.7 & 0.21 & 1.13 & 0.04 &~& -0.76 & 0.65 &  0.08 \\
            &      &      &      &      &      &~&  0.65 & 0.76 &  0.05 \\
Q2237+030   & 0.00 & 1.44 & 0.05 & 1.37 & 0.28 &~&  0.77 & 0.01 & -0.63 \\
            &      &      &      &      &      &~& -0.56 &-0.45 & -0.69 \\
H1413+117   & 0.21 & 0.45 & 0.10 & 1.21 & 0.97 &~& -0.63 & 0.57 &  0.52 \\
            &      &      &      &      &      &~&  0.77 & 0.48 &  0.41 \\
MG0414+0534 & 0.00 & 0.10 & 0.85 & 0.52 & 0.65 &~& -0.004&-0.01 &-0.999 \\
            &      &      &      &      &      &~& -0.68 &-0.73 & 0.01  
\end{tabular}
\caption{\footnotesize The $a_i$ coefficients are for the most well constrained
generalized magnification ratio on top and the least well constrained
ratio on the bottom ($a_4=-\sum_{i=1}^3 a_i$).  The third generalized
ratio is orthoginal to these two vectors.  The
numbering of images is in the same order as those in the observational
liturature.  $P(\chi^2)$ is the expected probability (rounded off) of
$\chi^2$ being smaller than the one measured with 10 degrees of
freedom.  Of the two models for B1422 the second one is a slightly
better fit to the image positions.
\label{table}}
\end{table}

% \begin{tabular}{lccccccc}
% & \multicolumn{5}{c}{Best--Fit Lens Models} &~&
% \multicolumn{1}{c}{Generalized Ratios} \\
% \multicolumn{1}{c}{Lens} & $P(\chi^2)$ & $b$ &
% $\gamma$ & $\epsilon^2$ & $n$ &~& 
% \begin{array}{ccc}
% $a_1$ & $a_2$ & $a_3$
% \end{array}\\
% B1422+231   & 0.00 & 0.91 & 0.25 & 1.58 & 0.32 &~& 
% \begin{array}{ccc}
%  -0.62 & ~0.78 & -0.14 \\
%  -0.62 & ~0.78 & -0.14 
% \end{array}\\
%             & 0.00 & 2.00 & 0.36 & 2.38 & 0.11 &~& 
% \begin{array}{ccc}
% -0.73 & ~0.69 & -0.03 \\
% -0.73 & ~0.69 & -0.03
% \end{array}\\
% PG1115+080  & 0.00 & 14.7 & 0.21 & 1.13 & 0.04 &~& 
% \begin{array}{ccc}
% -0.76 & ~0.65 & ~0.08 \\
% -0.76 & ~0.65 & ~0.08 \\
% \end{array}\\
% Q2237+030   & 0.00 & 1.44 & 0.05 & 1.37 & 0.28 &~&  
% \begin{array}{ccc}
% ~0.77 & ~0.01 & -0.63 \\
% ~0.77 & ~0.01 & -0.63
% \end{array} \\
% H1413+117   & 0.21 & 0.45 & 0.10 & 1.21 & 0.97 &~&  
% \begin{array}{ccc}
% ~0.32 & ~0.44 & -0.84 \\
% ~0.32 & ~0.44 & -0.84
% \end{array} \\
% MG0414+0534 & 0.00 & 0.10 & 0.85 & 0.52 & 0.65 &~& 
% \begin{array}{ccc}
% -0.004& -0.01 &  -0.999 \\
% -0.004& -0.01 &  -0.999 
% \end{array} 
% \end{tabular}
% \caption{\footnotesize The $a_i$ coefficients are for the most well constrained
% generalized magnification ratio ($a_4=-\sum_{i=1}^3 a_i$).  The
% numbering of images is in the same order as those in the observational
% liturature.  $P(\chi^2)$ is the expected probability (rounded off) of
% $\chi^2$ being smaller than the one measured with 10 degrees of
% freedom.  Of the two models for B1422 the second one is a slightly
% better fit to the image positions.
% \label{table}}
% \end{table}

\end{document}